\documentstyle[12pt]{article}
\begin{document}

\begin{center}
Physica Scripta, Vol. {\bf 59}, No. 4, pp. 251--256, 1999
\end{center}

\vspace{2cm}

\begin{center}
{\Large {\bf \ Kepler map }}\vspace{1cm} \\[\baselineskip] {B. Kaulakys$^{*}$
and G. Vilutis$^{\dagger }$}\\[\baselineskip] {Institute of Theoretical
Physics and Astronomy, A. Go\v stauto 12, 2600 Vilnius, Lithuania}

\vspace{2\baselineskip} {\bf Abstract }
\end{center}

{\small \ We present a consecutive derivation of mapping equations of motion
for the one-dimensional classical hydrogenic atom in a monochromatic field
of any frequency. We analyze this map in the case of high and low relative
frequency of the field and transition from regular to chaotic behavior. We
show that the map at aphelion is suitable for investigation of transition to
chaotic behavior also in the low frequency field and even for adiabatic
ionization when the strength of the external field is comparable with the
Coulomb field. Moreover, the approximate analytical criterion (taking into
account the electron's energy increase by the influence of the field) yields
a threshold field strength quite close to the numerical results. We reveal
that transition from adiabatic to chaotic ionization takes place when the
ratio of the field frequency to the electron Kepler frequency approximately
equals 0.1. For the dynamics and ionization in a very low frequency field
the Kepler map can be converted to a differential equation and solved
analytically. The threshold field of the adiabatic ionization obtained from
the map is only 1.5\% lower than the exact field strength of static field
ionization. }

\vspace{1cm}

PACS number(s): 05.45+b, 32.80.Rm, 42.50.Hz

\vspace{4cm}

---------------------------------------

$^{*}$ Electronic address: kaulakys@itpa.lt

$^{\dagger }$ Electronic address: gedas@itpa.lt

\newpage

{\bf 1. INTRODUCTION}

It is already the third decade when the highly excited hydrogen atom{\bf \ }
in a microwave field remains one of the simplest and most proper real system
for experimental and theoretical investigation of classical and quantum
chaos in the nonlinear systems strongly driven by external driving fields
(see reviews [1--5] and references herein). For theoretical analysis of
transition to stochastic behavior and ionization processes of atoms in
microwave fields approximate mapping equations of motion, rather than
differential equations, are most convenient.

Such a two-dimensional map (for the scaled energy of the one-dimensional
atom in a monochromatic field and for the relative phase of the field),
later called {\it Kepler map} [3,13], has been obtained in Refs. [6,7] after
an integration of equations of motion for one period of the intrinsic motion
of the electron between two subsequent passings of the aphelion, the largest
distance from the nucleus. This map greatly facilitates numerical
investigation of dynamics and ionization process and allows even an
analytical estimation of the threshold field strengths for the onset of
chaos, the diffusion coefficient of the electron in energy space and other
characteristics of the system [3--10]. Moreover, this map is closely related
to the expressions of quasiclassical dipole matrix elements for high atomic
states [11,12].

The Kepler map for relatively high frequencies of the field (recovered in
[3,13] and sometimes represented for the number of absorbed photons) is
relatively simple and is widely used for analysis of classical dynamics, as
well as after the quantization -- the quantum Kepler map for the quantum
dynamics [3--5,9,10,13]. In the region of relatively low frequency of the
field this map is sufficiently more complex, the threshold field strength
for transition to chaotic behavior and ionization is considerably higher
than that for transition to chaotic behavior in the medium and high
frequency fields.

Since derivation of the mapping equations of motion is based on the
classical perturbation theory and the electron's energy change due to
interaction with an external field during the period of motion in the
Coulomb field depends on the initial condition, i.e., on the integration
interval [6,7,11], complementary analysis of the applicability of the
'standard' Kepler map is necessary.

It should be noted, that in derivation of the Kepler map one integrates not
over the period of the external electromagnetic field but over the period of
the electron intrinsic motion in the Coulomb field [6,7]. This results in
some contradictions and difficulties. First, the period of integration and
the obtained map depend on the energy of the electron which complicates the
quantization problem of the Kepler map [3--7,13--17]. Second, the energy
change of the electron during the period of the classical intrinsic motion
due to interaction with the microwave field depends on the starting
conditions, i.e., on the integration interval [7,11]. This makes it
possibility to obtain another map, the Kepler map at perihelion, derived by
the integration between two subsequent passings of the perihelion [7] or
even a three-dimensional map for the two halves of the intrinsic period
[11]. These maps stronger reveal the resonance structure of chaotic dynamics
at low frequencies [6,7,11,15].

Moreover, Nauenberg [18] has presented a so-called {\it canonical Kepler map}
which agrees with the results of Refs. [6,7,11] if taken at perihelion but
not with the widely used Kepler map at aphelion. The expressions for the
variables of this canonical Kepler map are, however, sufficiently
complicated, not explicit and, therefore, they are not comfortable for
analytical and even numerical analysis. So, for this reason corroboration of
the standard Kepler map's applicability for the large range of parameters of
the problem is significant as well. Note additionally, that in derivation
and analysis of the maps in Ref. [7] some inaccuracy and misprints have
appeared.

All these aspects indicate the need of additional analysis of the mapping
equations of motion for a highly excited classical hydrogenic atom in a
monochromatic field. In addition, transition from adiabatic to chaotic
ionization mechanism in the a frequency field is of great interest (see,
e.g. [19] and references herein).

In this paper we present a consistent derivation of the mapping equations of
motion for a one-dimensional classical atom in a monochromatic field of any
frequency and analyze transition from regular to chaotic behavior and
ionization process. From the fulfilled analysis we conclude that the map at
aphelion and an approximate analytical criterion of the onset of chaos are
suitable also in the low frequency region, even for adiabatic ionization,
where the strength of the external field is comparable with the Coulomb
field. Moreover, in this case the map can be transformed to a differential
equation and solved analytically.

{\bf 2. MAPPING EQUATIONS\ OF\ MOTION }

The direct way of coupling the electromagnetic field to the electron
Hamiltonian is through the ${\bf A\cdot P}$ interaction, where ${\bf A}$ is
the vector potential of the field and ${\bf P}$ is the generalized momentum
of the electron. The Hamiltonian of the hydrogen atom in a linearly
polarized field $F\cos (\omega t+\vartheta )$, with $F$, $\omega $ and $
\vartheta $ being the field strength amplitude, field frequency and phase,
respectively, in atomic units is
\begin{equation}
H={\frac 12}\left( {\bf P}-{\frac{{\bf F}}\omega }\sin (\omega t+\vartheta
)\right) ^2-{\frac 1r}.
\end{equation}

The electron energy change due to interaction with the external field
follows from the Hamiltonian equations of motion [20]
\begin{equation}
\dot E=-{\bf \dot r\cdot F}\cos (\omega t+\vartheta ).
\end{equation}

Note, that Eq. (2) is exact if ${\bf \dot r}$ is obtained from equations of
motion including influence of the electromagnetic field. Using parametric
equations of motion in the Coulomb field we can calculate the change of the
electron's energy in the classical perturbation theory approximation
[6--8,11].

Measuring the time of the field action in the field periods one can
introduce the scale transformation [16,18] where the scaled field strength
and the scaled energy are $F_s=F/\omega ^{4/3}$ and $E_s=E/\omega ^{2/3}$,
respectively. However, it is convenient [6--8,17] to introduce the positive
scaled energy $\varepsilon =-2E_s$ and the relative field strength $
F_0=Fn_0^4=F_s/\varepsilon _0^2$, with $n_0$ being the initial effective
principle quantum number, $n_0=\left( -2E_0\right) ^{-1/2}$. The threshold
values of the relative field strength $F_0$ for the ionization onset depends
weaker upon the initial effective principle quantum number $n_0$ and the
relative frequency of the field $s_0=\omega n_0^3$ than the scaled field
strength $F_s$.

We restrict our subsequent consideration to the one-dimensional model, which
corresponds to states very extended along the electric field direction. Such
classical one-dimensional model was first considered in Refs. [21] for the
description of surface-state electrons, while a justification of the use of
one-dimensional-like states for periodically driven hydrogen atoms appeared
in [22]. Since that the one-dimensional model is widely used in theoretical
analysis [2--11,13--17].

For the derivation of a map describing the motion of an electron in the
superposition of the Coulomb and microwave fields we should integrate
dynamical equations over some characteristic period of the system. The
peculiarity of the system under consideration is that we are able to obtain
explicit expressions for the change of the electron energy only for halves
of the period and for the complete period, $T$ $=2\pi /(-2E)^{3/2}=2\pi
/\omega \varepsilon ^{3/2}$, of the intrinsic electron motion in the Coulomb
field [6--8,11] but not for the period of the external field.

Integration of Eq. (2) for motion between two subsequent passages at the
aphelion (where $\dot x=0$ and there is no energy exchange between the field
and the atom) results in the change of the electron's energy [6,7,11]
$$
\Delta E=-\left( \pi F/E\right) {\bf J}_s^{\prime }(s)\sin \vartheta .
$$
Here $s\equiv \varepsilon ^{-3/2}=\omega /(-2E)^{3/2}=\omega /\Omega $ is
the relative frequency of the field, i.e., the ratio of the field frequency $
\omega $ to the Kepler orbital frequency $\Omega =\left( -2E\right) ^{3/2}$,
and ${\bf J}_s^{\prime }(z)$ is the derivative of the Anger function with
respect to the argument $z$. The derivative of the Anger function
$$
{\bf J}_s^{\prime }(s)=\frac 1\pi \int\limits_0^\pi \sin \left[ s\left(
x-\sin x\right) \right] \sin x{\rm d}x
$$
is a very simple analytical function which can be approximated quite well by
some combination [7] of expansion in powers of $s$
$$
{\bf J}_s^{\prime }(s)=\frac{1+\left( 5/24\right) s^2}{2\pi \left(
1-s^2\right) }\sin \pi s,\quad s\leq 1
$$
and of the asymptotic form
$$
{\bf J}_s^{\prime }(s)=\frac b{s^{2/3}}-\frac a{5s^{4/3}}-\frac{\sin \pi s}{
4\pi s^2},\quad s\gg 1
$$
where
$$
a=\frac{2^{1/3}}{3^{2/3}\Gamma \left( 2/3\right) }\simeq 0.4473,\quad b=
\frac{2^{2/3}}{3^{1/3}\Gamma \left( 1/3\right) }\simeq 0.41085.
$$

Introducing the scaled energy $\varepsilon =-2E/\omega ^{2/3}$ and the
relative field strength $F_0=F/4E_0^2$ we have
$$
\Delta \varepsilon =-\pi F_0\varepsilon _0^2h\left( \varepsilon \right) \sin
\vartheta \eqno{(3)}
$$
where $\varepsilon _0=-2E_0/\omega ^{2/3}$ and
$$
h\left( \varepsilon \right) =\frac 4\varepsilon {\bf J}_s^{\prime }(s).
\eqno{(4)}
$$
The change of the field phase $\vartheta $ after the electron motion period
in the Coulomb field is
$$
\Delta \vartheta =2\pi \omega T=2\pi /\varepsilon ^{3/2}.\eqno{(5)}
$$
Defining the scaled energy and the phase before, $\varepsilon _j,\vartheta
_j $, and after, $\varepsilon _{j+1},\vartheta _{j+1}$, passages of the
electron of one intrinsic motion period we can introduce [23,24] a
generating function $G\left( \varepsilon _{j+1},\vartheta _j\right) $ of the
map determined as
$$
\varepsilon _j=\partial G/\partial \vartheta _j,\quad \vartheta
_{j+1}=\partial G/\partial \varepsilon _{j+1}.\eqno{(6)}
$$
In agreement with Eqs. (3) and (5) the generating function is (see also [3]
for analogy)
$$
G\left( \varepsilon _{j+1},\vartheta _j\right) =\varepsilon _{j+1}\vartheta
_j-4\pi \varepsilon _{j+1}^{-1/2}-\pi F_0\varepsilon _0^2h\left( \varepsilon
_{j+1}\right) \cos \vartheta _j\eqno{(7)}
$$
and according to Eqs. (6) it generates the map
$$
\left\{
\begin{array}{ll}
\varepsilon _{j+1}= & \varepsilon _j-\pi F_0\varepsilon _0^2h\left(
\varepsilon _{j+1}\right) \sin \vartheta _j, \\
\vartheta _{j+1}= & \vartheta _j+2\pi /\varepsilon _{j+1}^{3/2}-\pi
F_0\varepsilon _0^2\eta \left( \varepsilon _{j+1}\right) \cos \vartheta _j.
\end{array}
\right. \eqno{(8)}
$$
Here
$$
\eta \left( \varepsilon \right) =\frac{{\rm d}h\left( \varepsilon \right) }{
{\rm d}\varepsilon }.\eqno{(9)}
$$

Note that the map (8) can be derived also without introduction of the
generating function [6--8,11] but using the requirement of the
area-preserving of the map (8) defined as
$$
\frac{\partial \left( \varepsilon _{j+1},\vartheta _{j+1}\right) }{\partial
\left( \varepsilon _j,\vartheta _j\right) }=1.\eqno{(10)}
$$

It should also be noted that the map (14)-(19) in Ref. [7] is with the
positive signs of terms in the right-hand site of Eq. (8) containing the
field amplitudes $F_0$, i.e., it is derived for the reverse orientation of
the atom with respect to the field orientation. Also note that a function $
\sin \vartheta _k$ was inadvertently omitted from the right-hand side of Eq.
(15) in [7].

The map (8) is the general mapping form of the classical equations of motion
for the one-dimensional hydrogen atom in a microwave field derived in the
classical perturbation theory approximation. Some analytical and numerical
analysis of this map has already been done in Refs. [3,6--8]. Here we
analyze different special cases of the map (8).

{\bf 3. HIGH FREQUENCY\ LIMIT\ }

For relatively high frequencies of the field, $s\gg 1$ ($s\geq 2$),
theoretical analysis of the classical dynamics of the one-dimensional
hydrogen atom in a microwave field is relatively simple. That is why the
energy changes of the electron, $\left( E_{j+1}-E_j\right) $ and $\left(
\varepsilon _{j+1}-\varepsilon _j\right) $, do not depend on the initial
energy $\varepsilon _j$ and relative frequency $s\gg 1$. Indeed, using the
asymptotic form of the derivative of the Anger function, ${\bf J}_s^{\prime
}(s)=b/s^{2/3}$, we have $h\left( \varepsilon _{j+1}\right) =4b=const.$, $
\eta \left( \varepsilon _{j+1}\right) =0$ and, consequently, the following
map
$$
\left\{
\begin{array}{ll}
\varepsilon _{j+1}= & \varepsilon _j-4\pi bF_0\varepsilon _0^2\sin \vartheta
_j, \\
\vartheta _{j+1}= & \vartheta _j+2\pi /\varepsilon _{j+1}^{3/2}.
\end{array}
\right. \eqno{(11)}
$$

Note, that scaled classical dynamics according to maps (8) and (11) depends
only on single combination of the field parameters, i.e., on the scaled
field strength $F_s=F_0\varepsilon _0^2=F/\omega ^{4/3}$ (see also [16,17]).

By the standard [23,24] linearization procedure, $\varepsilon _j=\varepsilon
_0+\Delta \varepsilon _j$, in the vicinity of the integer relative frequency
(resonance), $s_0=\varepsilon _0^{-3/2}=m$ with $m$ integer, the map (11)
can be transformed to the standard (Chirikov) map
$$
I_{j+1}=I_j+K\sin \vartheta _j,
$$
$$
\vartheta _{j+1}=\vartheta _j+I_{j+1}.\eqno{(12)}
$$
Here $I_j=-3\pi \Delta \varepsilon _j/\varepsilon _0^{5/2}$ and $K=12\pi
^2bF_0/\sqrt{\varepsilon _0}$.

From the condition of the onset of classical chaos for the standard map, $
K\geq K_c\simeq 0.9816$ [1,23--25], we can, therefore, estimate the
threshold field strength for chaotization of dynamics and ionization of the
atom in the high frequency field
$$
F_0^c=K_c/\left( 12\pi ^2bs_0^{1/3}\right) \simeq 0.02s_0^{-1/3}.\eqno{(13)}
$$

Sometimes one writes the map (11) for a variable $N=-1/2n^2\omega $, which
change gives the number of absorbed photons [3,13],
$$
N_{j+1}=N_j+2\pi \left( F/\omega ^{5/3}\right) \sin \vartheta _j,
$$
$$
\vartheta _{j+1}=\vartheta _j+2\pi \omega \left( -2\omega N_{j+1}\right)
^{-3/2}.\eqno{(14)}
$$

We see that for such variables the dynamics of the system depends on {\it two
} parameters: on the ratio $F_q=F/\omega ^{5/3}$ (in Refs. [16,17] $
F_q=F/\omega ^{5/3}$ was called the {\it quantum scaled field strength}) and
on the field frequency $\omega $. Map (14) is, therefore, not the most
convenient one for analysis of the {\it classical} dynamics.

In general there are, however, no essential difficulties in the theoretical
analysis of classical nonlinear dynamics of the highly excited hydrogen atom
in the microwave field of relative frequency $s_0=\omega n_0^3\geq 0.5$ when
the field strength is lower or comparable with the threshold field strength
for the onset of classical chaos, i.e., if the microwave field is
considerably weaker than the characteristic Coulomb field. In such a case,
the energy change of the electron during the period of intrinsic motion is
relatively small and application of the classical perturbation theory for
derivation of the Kepler map (8) is sufficiently correct. Further analysis
of transition to chaotic behavior and of the ionization process can be based
on the map (8) and for $s_0\simeq 0.3\div 1.5$ results in the ''impressive
agreement'' [5] between measured ionization curves and those obtained from
the map (8) [5--10]. Even analytical estimation of the threshold field
strengths based on this map is rather proper [6--8].

Considerably more complicated is the analysis of transition to stochastic
motion and of ionization process in the region of low relative frequencies, $
s_0\leq 0.3$, [6--8,11].

{\bf 4. LOW\ FREQUENCY\ LIMIT }

For the low relative frequencies of the microwave field, $s\ll 1$, the map
(8) can be simplified as well. Using expansion of the function ${\bf J}
_s^{\prime }(s)$ in powers of $s$, ${\bf J}_s^{\prime }(s)\simeq s/2$, for $
s\ll 1$ we have according to Eqs. (4) and (9)
$$
h\left( \varepsilon _{j+1}\right) =2/\varepsilon _{j+1}^{5/2},
$$
$$
\eta \left( \varepsilon _{j+1}\right) =-5/\varepsilon _{j+1}^{7/2}.
\eqno{(15)}
$$
Consequently map (8) transforms to the form
$$
\left\{
\begin{array}{ll}
\varepsilon _{j+1}= & \varepsilon _j-2\pi F_0\left( \varepsilon
_0^2/\varepsilon _{j+1}^{5/2}\right) \sin \vartheta _j, \\
\vartheta _{j+1}= & \vartheta _j+2\pi /\varepsilon _{j+1}^{3/2}+5\pi
F_0\left( \varepsilon _0^2/\varepsilon _{j+1}^{7/2}\right) \cos \vartheta
_j.
\end{array}
\right. \eqno{(16)}
$$

This map is slightly more complicated than map (11) for high frequencies,
however, it can easily be analyzed numerically as well as analytically. Note
first of all, that the energy change of the electron during the period of
intrinsic motion (after one step of iteration), $\left| \varepsilon
_{j+1}-\varepsilon _j\right| ,$ is considerably smaller than the binding
energy of the electron $\varepsilon _j\simeq \varepsilon _0$ if the field
strength is lower or comparable with the threshold field strength, i.e., $
2\pi F_0\left( \varepsilon _0^2/\varepsilon _{j+1}^{5/2}\right) \simeq 2\pi
F_0\varepsilon _0^{-1/2}\ll \varepsilon _0$, or $2\pi F_0s_0\ll 1$ if $
F_0\leq F_0^{st}\simeq 0.13$ and $s_0\ll 1$. This indicates that the map
(16) is probably suitable for description of dynamics even in the low
frequency region where the field is relatively strong.

In Figs 1 and 2 the results of the numerical analysis of maps (8) and (16)
in the low frequency, $s\leq 1$, area are presented. We see that the
threshold ionization field calculated from the maps approaches the static
field ionization threshold $F_0^{st}\simeq 0.13$ when $s_0\rightarrow 0$.
This supports the presumption that the map (8) is valid even in the low
frequency limit where the strength of the driving field is of the order of
the Coulomb field.

4.1. Adiabatic ionization

For low frequencies, $2\pi s=2\pi /\varepsilon ^{3/2}\ll 1$, according to
the second equation of map (16) the change of the angle $\vartheta $ after
one step of iteration is small. As it was noticed above, the energy change
is also relatively small. Therefore, we can transform the difference
equations (16) to differential equations of the form
$$
\frac{{\rm d}\varepsilon }{{\rm d}j}=-\frac{2\pi \varepsilon _0^2F_0}{
\varepsilon ^{5/2}}\sin \vartheta ,
$$
$$
\frac{{\rm d}\vartheta }{{\rm d}j}=\frac{2\pi }{\varepsilon ^{3/2}}+\frac{
5\pi \varepsilon _0^2F_0}{\varepsilon ^{7/2}}\cos \vartheta .\eqno{(17)}
$$
Dividing second equation of the system (17) by the first one we obtain one
differential equation
$$
\frac{{\rm d}\left( \cos \vartheta \right) }{{\rm d}\varepsilon }=\frac
\varepsilon {\varepsilon _0^2F_0}+\frac{5\cos \vartheta }{2\varepsilon }.
\eqno{(18)}
$$
The analytical solution of Eq. (18) with the initial condition $\varepsilon
=\varepsilon _0$ when $\vartheta =\vartheta _0$ is
$$
\cos \vartheta =z^5\cos \vartheta _0-2z^4\left( 1-z\right) /F_0,\quad z=
\sqrt{\varepsilon /\varepsilon _0}.\eqno{(19)}
$$

Eq. (16) describes the motion of the system in $\varepsilon $ and $\vartheta
$ variables, i.e., represents the functional interdependence between two
dynamical variables.

Let us analyze Eqs. (18) and (19) in more detail. For relatively low values
of $F_0$, i.e., for $F_0<\frac 25z^4=\frac 25\left( \frac \varepsilon
{\varepsilon _0}\right) ^2$, the right-hand side of Eq. (18) is positive for
all phases $\vartheta $. Therefore, $\cos \vartheta $ and $\varepsilon $
decrease and increase simultaneously and, according to Eq. (16), there is a
motion in the whole interval $\left[ 0,2\pi \right] $ of the angle $
\vartheta .$ For $F_0>\frac 25z^4$, however, the increase of the angle $
\vartheta $ in the interval $0\div \pi $ goes to decrease of $\vartheta $ at
$\vartheta \simeq \pi $. This results in fast decrease of $\varepsilon $ and
to the ionization process (see also Fig. 1). It is easy to understand from
analysis of Eq. (19) that the minimal value of $F_0$ for such a motion
(resulting in ionization) corresponds to $\vartheta _0=0$ and $\vartheta
=\pi $. This value of $F_0$ is very close to the maximal value of $F_0$
resulting from motion in the whole interval $\left[ 0,2\pi \right] $ of $
\vartheta $, i.e., the maximum of the expression
$$
F_0=2z\left( 1-z\right) /\left( 1+z^5\right) .\eqno{(20)}
$$
This maximum is at $z=z_0$, where $z_0$ is a solution of the equation $
z^5+5z-4=0$, being $z_0\simeq 0.75193$. The critical value of the relative
field strength, therefore, is $F_0^c=2z_0^4/5=0.1279$ which is only 1.5\%\
lower than the adiabatic ionization threshold $F_0^{st}=2^{10}/\left( 3\pi
\right) ^4=0.1298$ [7,21]. According to our numerical analysis, if $s_0\leq
0.05$ the electron remains bounded and the dynamics is regular for $F_0\leq
0.13$ while ionization takes place for $F_0\geq 0.131$ (see also Fig. 1 and
2). These results are very close as well to analytical conclusions. Note,
that some decrease of the threshold field strength values $F_0^c$ with
decreasing of $s_0$ was observed for $s_0\leq 0.1$. Dynamics and classical
ionization at such frequencies are, however, essentially adiabatic.

4.2. Chaotic ionization

For higher relative frequencies, $s_0\geq 0.1$, ionization process is due to
chaotic dynamics of the highly excited electron of the hydrogenic atom in a
microwave field. There are different criterions for estimation of the
parameters when the dynamics of the nonlinear system becomes chaotic. For
analysis of transition to chaotic behavior of the motion described by maps
(8), (11) and (16) the most proper, to our mind, is the criterion related
with the randomization of the phases [24]
$$
K=\max \left| \frac{\delta \vartheta _{j+1}}{\delta \vartheta _j}-1\right|
\geq 1.\eqno{(21)}
$$
Here $\max $ means the maximum with respect to the phase $\vartheta _j$ and
variation of the phase $\vartheta _{j+1}$ with respect to the phase $
\vartheta _j$ means the full variation including dependence of $\vartheta
_{j+1}$ on $\vartheta _j$ through the variable $\varepsilon _{j+1}$ in Eqs.
(8), (11) and (16).

Applying criterion (21) to the general map (8) we obtain the threshold field
strength
$$
F_0^c=\frac{\varepsilon ^{7/2}}{12\pi ^2\varepsilon _0^2{\bf J}_s^{\prime
}(s)}.\eqno{(22)}
$$
If $\varepsilon \simeq \varepsilon _0$ Eq. (22) yields the result
$$
F_0^c=\left( 12\pi ^2s{\bf J}_s^{\prime }(s)\right) ^{-1}\eqno{(23)}
$$
which for $s\gg 1$ coincides with Eq. (13).

For more precise evaluation of the critical field strengths we should take
into account the change (increase) of the electron's energy due to the
influence of the electromagnetic field. For higher relative frequency $s$ or
lower scaled energy $\varepsilon _j$ the threshold ionization field is
lower. Therefore, if the scaled energy $\varepsilon _j$ decreases as a
result of relatively regular dynamics in a not very strong microwave field,
then the lower field strength is sufficient for transition to chaotic
dynamics. For high frequencies such change of the energy is relatively
small. Nevertheless it reveals some resonance structure in the field-atom
interaction. In the low frequency limit the energy change is more essential.
Now consider it in detail.

As it was shown above, maximal decrease of the scaled energy $\varepsilon _j$
is for the angle $\vartheta _j\simeq \pi $ and it can be evaluated from Eq.
(20). Taking this into account we have from Eq. (16) according to criterion
(21) the expression for the threshold relative field strength
$$
F_0^c=\frac{\varepsilon ^5}{6\pi ^2\varepsilon _0^2}=\frac{z_c^{10}}{6\pi
^2s_0^2}\eqno{(24)}
$$
where $z_c$ is the solution of Eq. (20) with $F_0=F_0^c$. Eq. (20) can be
solved approximately expanding $z_c^{10}$ in powers of $F_0^c$. The result
of such an expansion is
$$
z_c^{10}\simeq 1-10F_0^c+30\left( F_0^c\right) ^2-73.6\left( F_0^c\right) ^3
\eqno{(25)}
$$
where the last term in the right-hand side of Eq. (25) is from the
requirement of the exact maximal value $z_c=z_0=0.75193$ for the static
threshold field strength $F_0^c=0.1279$.

For evaluation of the threshold field for transition to chaotic behavior in
the low frequency field we should thus solve the system of equations (24)
and (25). For $0.09\leq s_0\leq 0.5$ expressions (24) and (25) give an
ionization threshold field very close to the numerical results (see Fig. 2).
For frequencies lower than $s_0\simeq 0.09$ ionization is adiabatic, because
for so low frequencies the adiabatic ionization threshold field, $
F_0^c=2z_0^4/5=0.1279$, is lower than the phase randomization field
evaluated according to Eqs. (24) and (25). The adiabatic ionization,
therefore, occurs in such a case earlier than the chaotization of the
dynamics. Note, that numerical results reveal transition from adiabatic to
chaotic ionization at relative frequency $s_0\simeq 0.1$ (scaled energy $
\varepsilon _0\simeq 4.3$) as well.

At higher frequencies ionization is due to chaotic dynamics while transition
to chaotic behavior can be evaluated from the approximate criterion (21)
taking into account the electron's energy change by the influence of the
electromagnetic field. For frequencies higher than $s_0\simeq 0.5$ we should
use a more exact expression than ${\bf J}_s^{\prime }(s)\simeq s/2$ for the
derivative of the Anger function, i.e., Eq. (22).

{\bf 5. CONCLUDING REMARKS }

From the analysis given in this study we can conclude that the map at
aphelion (8) is suitable for investigation of regular and stochastic
classical dynamics, transition to chaotic behavior and ionization of Rydberg
atoms in high, medium and low frequency fields, even for adiabatic
ionization when the strength of the external field is comparable with the
averaged Coulomb field. For such a purpose it is unnecessary to use the map
at perihelion, the map for two halves of the intrinsic period or the
canonical Kepler map [7,11,18]. Moreover, the approximate criterion (21) for
transition to chaotic behavior yields a threshold field strength very close
to the numerical results if we take into account increase of the electron
energy by influence of the electromagnetic field. Transition from adiabatic
to chaotic ionization of the classical hydrogenic atom in a monochromatic
field takes place at a relative field frequency $s_0\simeq 0.1$.

Furthermore, the Kepler map and some generalizations of it (for two- and
multi-frequency [9,10,26] or some other fields, e.g., circular polarized
microwave field, for three-dimensional atoms and other modifications) are
and may be more widely used for analysis of different effects of classical
and quantum chaos in driven nonlinear systems [19,27,28]. Note also the
attempts to derive and use similar maps in astronomy for analysis of chaotic
dynamics of comets and other astronomical bodies [29]. It turns out,
however, that in such a case, generalization of the Kepler map for
nonharmonic perturbations and for motion in three-dimensional space, is
necessary.

{\bf ACKNOWLEDGMENT }

{\small The research described in this publication was made possible in part
by support of the Alexander von Humboldt Foundation.}

\vskip\baselineskip

{\bf References }

\begin{enumerate}
\item  Delone, N. B., Krainov, V. P. and Shepelyansky, D. L., Usp. Fiz. Nauk
{\bf 140}, 355 (1983) [Sov. Phys.-Usp. {\bf 26}, 551 (1983)].

\item  Casati, G., Chirikov, B. V., Shepelyansky, D. L. and Guarneri, I.,
Phys. Rep. {\bf 154}, 77 (1987).

\item  Casati, G., Guarneri, I. and Shepelyansky, D. L., IEEE J. Quantum
Electron. {\bf 24}, 1420 (1988).

\item  Jensen, R. V., Susskind, S. M. and Sanders, M. M., Phys. Rep. {\bf
201 } , 1 (1991).

\item  Koch, P. M., in ''Chaos and Quantum Chaos'', edited by Heiss, W.
(Lecture Notes in Physics Vol. {\bf 411,} Springer-Verlag, Berlin, 1992), p.
167; Koch, P. M. and van Leeuwen, K. A. H., Phys. Rep. {\bf 255}, 289 (1995).

\item  Gontis, V. and Kaulakys, B., Deposited in VINITI as No.5087-V86
(1986) and Lit. Fiz. Sb. {\bf 27}, 368 (1987) [Sov Phys.-Collect. {\bf 27},
111 (1987)].

\item  Gontis, V. and Kaulakys, B., J. Phys. B: At. Mol. Opt. Phys. {\bf 20}
, 5051 (1987).

\item  Kaulakys, B. and Vilutis, G., in ''Chaos - The Interplay between
Stochastic and Deterministic Behaviour''{\it ,} edited by Garbaczewski, P.,
Wolf, M. and Weron, A. (Lecture Notes in Physics Vol. {\bf 457,}
Springer-Verlag, Berlin, 1995), p. 445; chao-dyn/9503011.

\item  Moorman, L., Galvez, E. J., Sauer, B. E., Mortazawi-M., A., van
Leeuwen, K. A. H., v.Oppen, G. and Koch, P. M., Phys. Rev. Lett. {\bf 61},
771 (1988); Galvez, E. J., Sauer, B. E., Moorman, L., Koch, P. M. and
Richards, D., Phys. Rev. Lett. {\bf 61}, 2011 (1988); Koch, P. M., Moorman,
L., Sauer, B. E., Galvez, E. J. and Leeuwe, K. A. H., Phys. Scripta {\bf T26}
, 51 (1989); Haffmans, A., Bl\"umel, R., Koch, P. M. and Sirko, L., Phys.
Rev. Lett. {\bf 73}, 248 (1994).

\item  Siko, L. and Koch, P. M., Appl. Phys. B {\bf 60}, S195 (1995); Koch,
P. M., Physica D {\bf 83}, 178 (1995).

\item  Kaulakys, B., J. Phys. B: At. Mol. Opt. Phys. {\bf 24}, 571 (1991).

\item  Kaulakys, B., J. Phys. B: At. Mol. Opt. Phys. {\bf 28}, 4963 (1995);
physics/9610018.

\item  Casati, G., Guarneri, I. and Shepelyansky, D. L., Phys. Rev. A {\bf
36 }, 3501 (1987).

\item  Graham, R., Europhys. Lett. {\bf 7}, 671 (1988); Leopold, J. G. and
Richards, D., J. Phys. B: At. Mol. Opt. Phys. {\bf 23}, 2911 (1990).

\item  Gontis, V. and Kaulakys, B., Lit. Fiz. Sb. {\bf 28}, 671 (1988) [Sov.
Phys. - Collec. {\bf 28} (6), 1 (1988)]; Gontis V. and Kaulakys, B., Lit.
Fiz. Sb. {\bf 31}, 128 (1991) [Lithuanian Phys. J. (AllertonPress, Inc.)
{\bf 31} (2), 75 (1991)]

\item  Kaulakys, B., Gontis, V., Hermann, G. and Scharmann, A., Phys. Lett.
A {\bf 159}, 261 (1991); Kaulakys, B., Acta Phys. Pol. B {\bf 23}, 313
(1992).

\item  Kaulakys, B., Gontis, V. and Vilutis, G., Lith. Phys. J. (Allerton
Press, Inc.) {\bf 33}, 290 (1993); Kaulakys, B. and Vilutis, G., in {\it AIP
Conf. Proc.} (AIP, New York) {\bf 329}, 389 (1995); quant-ph/9504007.

\item  Nauenberg, M., Europhys. Lett. {\bf 13}, 611 (1990).

\item  Bl\"umel, R., Phys. Rev. A {\bf 49,} 4787 (1994); Sundaram, B. and
Jensen, R. V., Phys. Rev. A {\bf 51,} 4018 (1995); Koch, P. M., Acta Phys.
Pol. A {\bf 93}, 105 (1998).

\item  Landau, L. D. and Lifshitz, E. M., ''Classical Field Theory''
(Pergamon, New York, 1975).

\item  Jensen, R. V., Phys. Rev. Lett. {\bf 49}, 1365 (1982); Phys. Rev. A
{\bf 30}, 386 (1984).

\item  Shepelyansky, D. L., in {\it Proc. Intern. Conf. on Quantum Chaos},
Como, 1983 (Plenum, New York, 1995), p. 187.

\item  Lichtenberg, A. J. and Lieberman, M. A., ''Regular and Stochastic
Motion'' ( Springer-Verlag, New York, 1983 and 1992).

\item  Zaslavskii, G. M., ''Stochastic Behavior of Dynamical Systems''
(Nauka, Moscow, 1984; Harwood, New York, 1985).

\item  Jensen, R. V., Am. Scient. {\bf 75}, 168 (1987).

\item  Howard, J. E., Phys. Lett. A {\bf 156}, 286 (1991); Kaulakys, B.,
Grauzhinis, D. and Vilutis, G., Europhys. Lett. {\bf 43}, 123 (1998);
physics/9808048.

\item  Casati, G., Guarneri, I. and Mantica, G., Phys. Rev. A {\bf 50}, 5018
(1994); Buchleitner, A., Delande, D., Zakrzewski, J., Mantenga, R. N.,
Arndt, M. and Walther, H., Phys. Rev. Lett. {\bf 75}, 3818 (1995); Wojcik,
M., Zakrzewski, J. and Rzazewski, K., Phys. Rev. A {\bf 52}, 2523 (1995);
Sanders, M. M. and Jensen, R. V., Am. J. Phys. {\bf 64}, 21 and 1013 (1996);
Sacha, K. and Zakrzewski, J., Phys. Rev. A {\bf 55}, 568 (1997).

\item  Jensen, R. V., Nature (London) {\bf 355}, 311 (1992); Sirko, L.,
Haffmans, A., Bellermann, M. R. W. and Koch, P. M., Europhys. Lett. {\bf 33}
, 181 (1996); Brenner, N. and Fishman, S., Phys. Rev. Lett. {\bf 77}, 3763
(1996) and J. Phys. A {\bf 29}, 7199 (1996); Benvenuto, F., Casati, G. and
Shepelyansky, D. L., Phys. Rev. A {\bf 55}, 1732 (1997); Buchleitner, A. and
Delande, D., Phys. Rev. A {\bf 55}, 1585 (1997).

\item  Petrosky, T. Y., Phys. Lett. A {\bf 117}, 328, (1986); Sagdeev, R. Z.
and Zaslavsky, G. M., Nouvo. Cim. B {\bf 97}, 119 (1987); Chirikov, B. V.
and Vecheslavov, V. V., Astron. Astrophys. {\bf 221}, 146 (1989); Torbett,
V. M. and Smoluchowski, R., Nature (London) {\bf 345}, 49 (1990); Milani, A.
and Nobili, A. M., Nature (London) {\bf 357}, 569 (1992); Chicone, C. and
Retzloff, D. G., J. Math. Phys. {\bf 37}, 3997 (1996).
\end{enumerate}

\begin{center}
\vspace{3cm} {\bf Caption for the figures} {\Large \ }\vspace{1.5cm}
\end{center}

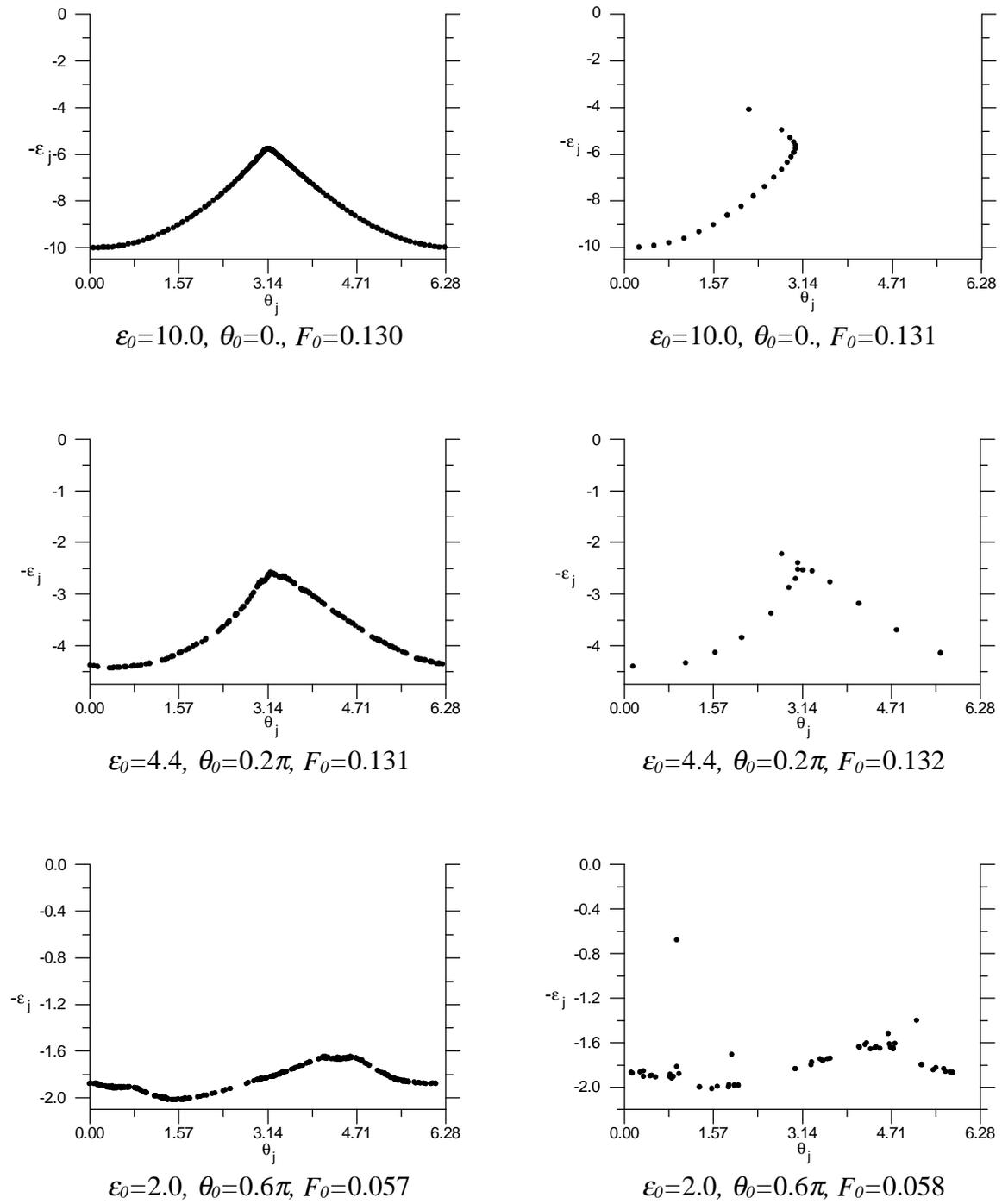
Fig. 1. Trajectories of the map (8) for different initial conditions, $
\varepsilon _0,\theta _0$, and different relative field strength $F_0$. The
pictures in the left-hand side correspond to the regular quasiperiodic
motion while those in the right-hand side represent ionization process for a
little stronger field. At $\varepsilon _0\simeq 4.3$, i.e., $s_0\simeq 0.11$
a transition from the adiabatic to the chaotic ionization mechanism takes
place. \vskip 2\normalbaselineskip

Fig. 2. Relative threshold field strength for the onset of ionization from
numerical analysis of the maps (8) and (16) and according to the approximate
criterion (24) - (25).

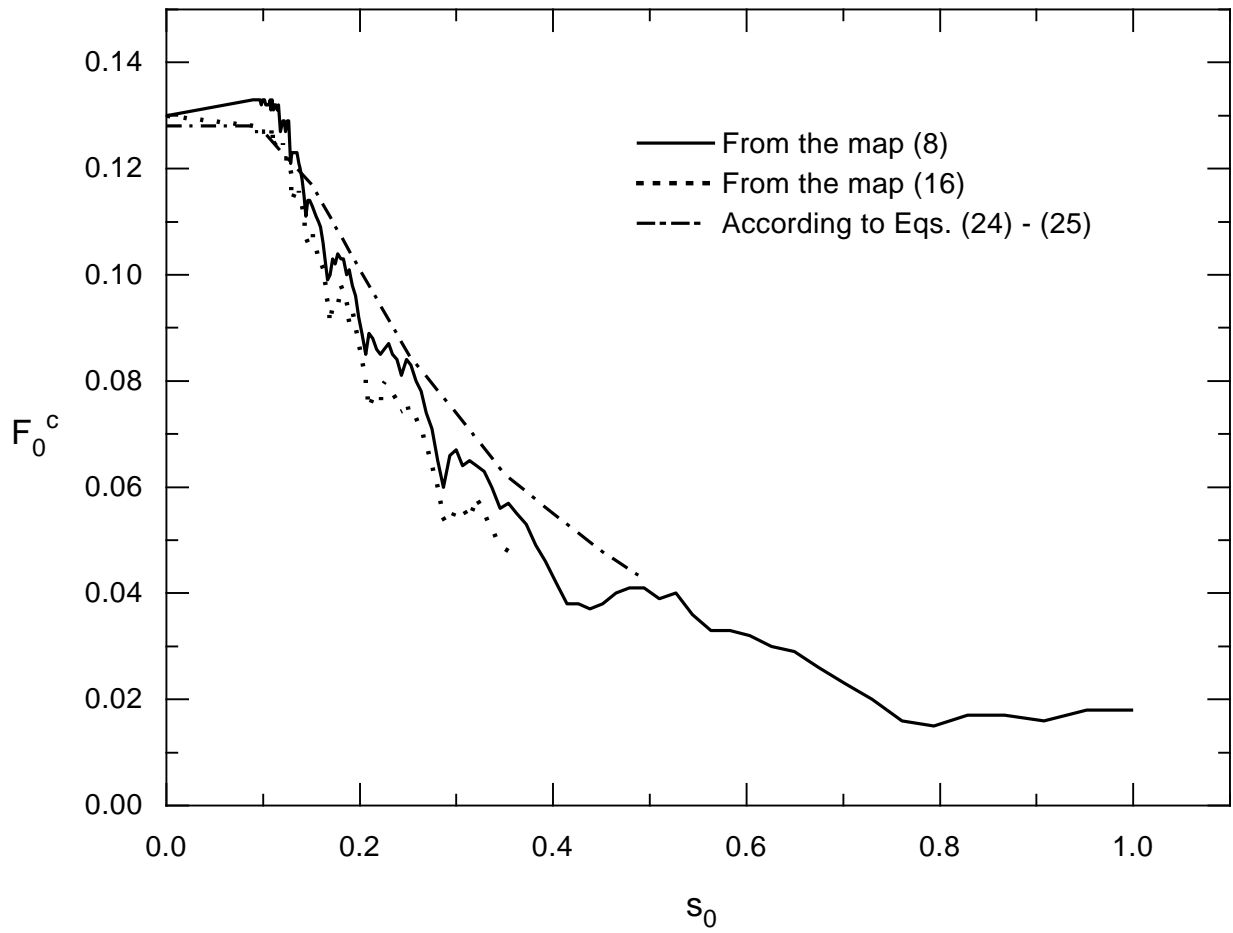
\end{document}